\documentclass[review,12pt]{elsarticle}

\usepackage[noend]{algpseudocode}
\usepackage{algorithm}

\usepackage[nointegrals]{wasysym} 

\algrenewcommand\algorithmicrequire{\textbf{Input:}}
\algrenewcommand\algorithmicensure{\textbf{Output:}}
\algrenewcommand\algorithmicif{\textbf{if}}
\algrenewcommand\algorithmicthen{}
\algrenewcommand\algorithmicelse{\textbf{else}}
\algrenewcommand\algorithmicfor{\textbf{for}}
\algrenewcommand\algorithmicdo{}
\algrenewcommand\algorithmicwhile{\textbf{while}}
\algrenewcommand\algorithmicreturn{\textbf{return}}
\algrenewcommand\algorithmicprocedure{\textbf{procedure}}

\usepackage{microtype}

\usepackage{siunitx}
\DeclareSIUnit{\litre}{l}

\usepackage[american]{babel}

\usepackage{booktabs}

\usepackage{geometry}
\usepackage{amsmath}
\usepackage{amssymb}
\usepackage{enumerate}
\usepackage{hyperref}

\usepackage{chemformula}

\usepackage{caption}
\usepackage[labelformat=simple]{subcaption}
\usepackage{placeins}

\begin{document}
 
\begin{frontmatter}
     
 \title{Complexity and nonlinearity of colloid electrical transducers}
     
 \author[1]{Raphael Fortulan\footnote{Corresponding author: raphael.vicentefortulan@uwe.ac.uk}}
 \author[1]{Noushin Raeisi Kheirabadi}
 \author[2,1]{Alessandro Chiolerio}
 \author[1]{Andrew Adamatzky}
     
 \affiliation[1]{Unconventional Computing Laboratory, UWE, Bristol, UK}
 \affiliation[2]{Bioinspired Soft Robotics Laboratory, Istituto Italiano di Tecnologia, Via Morego 30, 16165 Genova, Italy}

 \begin{abstract}
           
 This work explores the complexity and nonlinearity of seven different colloidal suspensions---Au, ferrofluid, \ch{TiO2}, ZnO, \ch{g-C3N4}, MXene, and PEDOT:PSS---when electrically stimulated with fractal, chaotic, and random binary signals. The recorded electrical responses were analyzed using entropy, file compression, fractal dimension, and Fisher information measures to quantify complexity. The nonlinearity introduced by each colloid was evaluated by the deviation of the output from the best-fit hyperplane of the input-output mapping. The results showed that \ch{TiO2} was the most complex colloid across all inputs, exhibiting high entropy, poor compressibility, and an unpredictable response pattern. The colloids also exhibited significant nonlinearity, making them promising candidates for reservoir computation, where the mapping of inputs into high-dimensional nonlinear states is advantageous. This study provides insight into the dynamics of colloids and their potential for unconventional computational applications that exploit their inherent complexity and nonlinearity, and it provides a rapid method for assessing the suitability of a particular material for use as a computational substrate before others.
           
 \end{abstract}

% \begin{keyword}
           
           
 %Colloids\sep Complexity Analysis\sep Nonlinearity\sep Unconventional Computing
           
 %\end{keyword}
               
\end{frontmatter}
 
 
\section{Introduction}
 Colloids, dispersed nanoparticles in a liquid, exhibit rich dynamics and physical phenomena arising from the interactions between the particles and the suspending medium. The application of external fields, such as electric or magnetic fields, can further influence the behavior of these systems, leading to the emergence of complex patterns and nonlinear responses. Harnessing these dynamics holds promise for unconventional computing paradigms, including reservoir and \textit{in-materia} computing~\cite{stepney2019co,miller2018materio,miller2019alchemy,lukovsevivcius2009reservoir,kheirabadi2023neuromorphic}.
 
The notion of computing with liquid media, also known as liquid computing, dates back over 120 years to early hydraulic algebraic machines~ \cite{adamatzky2019brief}. Since then, various approaches like fluid maze solvers, droplet logics, and colloid-based information processors have been explored, taking advantage of unique properties such as reconfigurability, scalability, potential low energy consumption, and inherent parallelism~\cite{fortulan2023reservoir,robertsLogicalCircuitsColloids2024}. In addition, previous works have demonstrated that colloid systems can implement synaptic learning and pattern recognition \cite{chiolerio2020liquid,crepaldi2023experimental}.

In this work, we explore the complexity and nonlinearity of seven distinct colloidal suspensions---Au, ferrofluid, \ch{TiO2}, ZnO, \ch{g-C3N4}, MXene, and PEDOT:PSS---when subjected to fractal, chaotic, and random binary electrical signals. By analyzing the recorded electrical responses using various complexity measures and quantifying the degree of nonlinearity introduced by the colloids, we aim to identify the most promising colloidal systems for potential use in reservoir computing architectures. The findings from this study provide valuable insights into the rich dynamical behaviors exhibited by colloids under electrical stimulation and their suitability for unconventional computing applications.
\section{Materials Fabrication}
A total of seven colloids were prepared to study their complexity and nonlinearity. Each colloid preparation method is described as follows:

\textbf{Au:} A colloid solution of Au nanoparticles dispersed in \ch{H2O} with a particle size of \SI{14}{\nano\metre} and a concentration of \SI{0.1}{\milli\gram\per\milli\litre} was provided by PlasmaChem GmbH (Germany).

\textbf{Ferrofluid:} A ferrofluid solution containing \ch{Fe3O4} nanoparticles dispersed in \ch{H2O} at a concentration of 30~wt\% was supplied by PlasmaChem GmbH (Germany).

\textbf{\ch{TiO2}:} A stable colloid of \ch{TiO2} nanoparticles (30-50 nm), anatase phase, in \ch{H2O} 40~wt\% was purchased from US Research Nanomaterials (USA).

\textbf{ZnO:} Sodium dodecyl sulfate (SDS, Merck) was added to deionized water to create a surfactant solution with a concentration of 0.22~wt\%. The solution was then stirred continuously to homogenize. In sequence, \SI{1}{\milli\gram} ZnO nanoparticles (US Research Nanomaterials, USA) were added to dimethyl sulfoxide (DMSO, US Research Nanomaterials, USA) while stirring continuously. The mixture was then stirred while \SI{2}{\milli\litre} of SDS solution and \SI{1}{\milli\litre} of \SI{10}{\mol\per\litre} NaOH were added. The resultant dispersion concentration was maintained constant at \SI{0.1}{\milli\gram\per\milli\litre}. The stirring operation was then continued for a few more hours to achieve a homogeneous dispersion of ZnO~\cite{kheirabadi2023learning, raeisi2024pavlovian}.

\textbf{\ch{g-C3N4}:} To synthesize \ch{g-C3N4} nanosheets, urea was first finely ground in an agate mortar and pestle. Then \SI{25}{\gram} of the powder was placed in an alumina crucible, sealed, and heated up in a muffle furnace up to \SI{550}{\celsius} at a rate of \SI{25}{\celsius\per\minute}. This temperature was maintained for 3 h to attain a yellow \ch{g-C3N4}. The acquired \ch{g-C3N4} was then ground into powder using an agate mortar and pestle~\cite{raeisi2020z}.
To achieve a suspension of \ch{g-C3N4} in \ch{H2O} with a concentration of \SI{1}{\milli\gram\per\milli\litre}, \SI{10}{\milli\gram} of \ch{g-C3N4} was mixed with \SI{10}{\milli\litre} deionized water~\cite{kheirabadi2023sustainable}.
 
\textbf{MXene:} In the first stage of MXene synthesis, a sacrificial etching process was employed. A solution of lithium fluoride (LiF, \SI{0.8}{\gram}) in hydrochloric acid (HCl, \SI{10}{\milli\litre}, \SI{9}{\mol\per\litre}) was stirred for 5 min, facilitating the in situ production of a small amount of hydrofluoric acid (HF). The MAX phase powder (\SI{0.5}{\gram} \ch{Ti3AlC2}) was then gradually added to the etching solution under continuous stirring. The sealed container was then placed on a magnetic stirrer at \SI{35}{\celsius} and 1000~rpm for 24 h. This allowed selective etching of the aluminum layers within the MAX phase by the HF~\cite{alhabeb2017guidelines}.

Following the etching step, the resultant mixture was centrifuged (3500~rpm, 5 min) with deionized water. This procedure eliminated the etching residue and excess acid. The centrifugation cycles were repeated until the supernatant reached a neutral pH ($\sim$6). According to reports, the remaining supernatant is a green, stable colloidal solution of MXene flakes dispersed in water at a low concentration (\SI{1}{\milli\gram\per\milli\litre}), indicating that delamination has been achieved~\cite{hantanasirisakul2019effects}. The remaining, darker part, which most likely contained larger MXene nanosheets, was sonicated with \SI{50}{\milli\litre} of deionized water for 1 h to encourage further delamination. The sonicated mixture was then recentrifuged to separate the water. Sediment containing MXene flakes was collected and dried in a vacuum oven at \SI{120}{\celsius}. The dried MXene powder was finally stored in a sealed container for future use~\cite{dong2018metallic, jiang2019all}. To prepare a suspension of MXene in dimethylformamide (DMF) at a concentration of \SI{1}{\milli\gram\per\milli\litre}, \SI{10}{\milli\gram} of as-synthesized MXene was mixed with \SI{10}{\milli\litre} DMF under stirring 1000~rpm for 24 h.

\textbf{PEDOT:PSS:} A high-conductivity grade solution of PEDOT:PSS, with a concentration of 3-4~wt\% in \ch{H2O}, was purchased from Merck.
Before conducting the studies, the suspensions were subjected to an ultrasonic bath (\SI{40}{\kilo\hertz}, DK Sonic Ultrasonic, UK) for 30 min to guarantee uniform dispersion of the particles.
\section{Experimental setup}
The electrical signals were generated using an Arduino Uno R4 board (Arduino, Italy) connected via I2C to two MCP4728 12-bit quad-channel digital-to-analog converters (DACs, Microchip Technology, USA), using a 74HC595 multiplexer (Texas Instruments, USA). The output channels of the DACs were connected to amplifier/level shift circuits based on LM258AP op-amps (Texas Instruments, USA) to generate bipolar outputs of $\pm$\SI{5}{\volt}. Both positive and negative voltage rails were supplied by an IPS 4303 Laboratory DC Power Supply (ISO-TECH, Taiwan). The outputs were then electrically connected to the colloids using \diameter \SI{10}{\micro\metre}  platinum/iridium-coated stainless steel probes to be used as inputs. An output probe was inserted into the colloid, and its voltage was measured using a 24-bit data logger from Pico Technology, UK.

To investigate the complexity and nonlinearity of all suspensions, three distinct signal patterns were programmed into the microcontroller: a random binary signal, a fractal curve, and a chaotic signal. The random binary signal, composed of a sequence of 1s and 0s generated randomly, aimed to probe the response of colloids to stochastic perturbations. The fractal curve, exhibiting self-similar patterns at different scales, was chosen to explore the hierarchical nature of colloidal structures. Meanwhile, the chaotic signal, characterized by its sensitive dependence on initial conditions, was intended to study the potential chaotic dynamics within the colloids.

By subjecting colloids to various stimuli, we aimed to understand their capabilities for advanced sensing and complex computation, as well as their usage in unconventional computing and reservoir computing. Each signal is described in more detail in the following sections.
\subsection{Koch snowflake}
Koch snowflake (or Koch curve) is one of the earliest and most famous fractal curves~\cite {ungarKochCurveGeometric2007}. In this work, the curve was generated using Algorithm S1, as shown in the supporting information. The curve represented by two signals $x$ and $y$ was then applied to the colloid using two probes.
 
\subsection{Lorenz system}
The Lorenz system is a chaotic dynamical system and is modeled by the following set of differential equations
\begin{align}
 \dot{x} & =\sigma(y-x), \\
 \dot{y} & =x(\rho-z),   \\
 \dot{z} & =xy-\beta z,  
\end{align}
where $\sigma$, $\rho$, and $\beta$ are parameters. In this work, the dynamical system was implemented on the microcontroller using the 4th-order Runge-Kutta method~\cite{burden2010numerical} with a time step of 100 ms. The choice of parameters was $\sigma=10$, $\rho=28$, and $\beta=8/3$, and the initial condition was set to $[1.2, 1.4, 1.6]^\intercal$.%The phase portrait and the solution in time of each state variable are presented in Figs.~\ref{fig:lorenz}(a) and (b), respectively.
%\begin{figure}[ht]
 %\centering
%    \includegraphics[width=.75\textwidth]{Figs/PDF/Plot_Lorenz.pdf}
 %\caption{(a) Phase portrait and (b) solution in time for each state variable for $\sigma=10$, $\rho=28$, and $\beta=8/3$.}
 %\label{fig:lorenz}
%\end{figure}
The $x, y,$ and $z$ states were then applied to the colloid using three probes. 
\subsection{Random Bombardment}
For the random bombardment, a random number between 0 and 255 (which corresponds to 8 bits) was selected every 500 ms, converted to its binary representation, and applied to the material using 8 probes. Bipolar logic was used, with +\SI{5}{\volt} representing the logical 1 and -\SI{5}{\volt} representing the logical 0.
\section{Black-Box Complexity}
We picture that the colloid is a black box that hides a function $f: V \rightarrow W$ inside. Assuming that no structural information is known about $f$, the only way to learn about it is by evaluating the function values of $f(x)$ at potential values $x \in V$. In addition, it is also possible to assume that the evaluation is performed by some \textit{oracle}, from which $f(x)$ is queried, which represents the signal generator and acquisition system. The complexity of $f$ (and by extension of the colloid) can then be inferred from a set of discrete tuples ${(x_0, f(x_0)), (x_1, f(x_1)), \ldots, (x_{N - 1}, f(x_{N-1}))}$.

For chaotic and dynamical systems—systems characterized by sensitive dependence on initial conditions and nonlinear dynamics—the definition of complexity for a time series can be split into two categories: (1) regularity and (2) predictability~\cite{bolltRegularizedForecastingChaotic2017, sobottkaPeriodicityPredictabilityChaotic2006}. Regularity refers to the number of repetitive patterns in the signal or the smoothness of the signal. Predictability, on the other hand, refers to the correlation present in the signal, i.e., how much information it carries about itself.
In this work, several complexity measures are used to characterize the output time series data of the colloid. Fractal dimension quantifies the complexity repeated in the signal as the ratio of the log number of boxes to the log of the inverse box length, $\log(N)/\log(1/L)$ \cite{wuEffectiveMethodCompute2020}. Shannon's entropy, $H(X) = -\sum^{N-1}_{i=0} P(x_i)\log_2 P(x_i)$ \cite{shannonMathematicalTheoryCommunication1948}, measures the minimum data storage requirements. Fisher information quantifies the amount of self-information carried by the system, which is anti-correlated with complexity. Compression algorithms like Gzip, bzip2, and LZMA2 assess compressibility, with more complex or random signals being less compressible. Further details on these complexity measures are available in the supporting information.

The computed entropy, gzip size, bzip size, XZ size, Fisher information, and fractal dimension of the recorded electrical signals from the colloids under fractal, chaotic, and random stimuli are presented in Tabs.~\ref{tab:snowflake} to~\ref{tab:Bombardment}, respectively. The most complex colloidal suspension was chosen as the one exhibiting the highest complexity indices and was identified as the most complex. Subsequent selections proceeded accordingly.
\begin{table}[ht]
 \centering
 \caption{The calculated entropy, gzip size, bzip size, XZ size, Fisher information, and fractal dimension of the measured electrical signal of the colloids for a fractal-like signal input}
 \label{tab:snowflake}
 \resizebox{\textwidth}{!}{
 \begin{tabular}{lSSSSSSS}
 \toprule
 \textbf{Colloid} & {\textbf{Entropy}} & {\textbf{gzip Size}} & {\textbf{bzip2 Size}} & {\textbf{XZ Size}} & {\textbf{FI}} & {\textbf{FD}} \\ \midrule
 Au               & 10.5               & 5.41                 & 4.74                  & 4.06               & 7.63E-01      & 1.02          \\
 Ferrofluid       & 10.5               & 3.53                 & 2.99                  & 2.71               & 9.00E-01      & 1.06          \\
 \ch{g-C3N4}      & 10.5               & 5.25                 & 4.65                  & 4.13               & 6.25E-01      & 1.03          \\
 MXene            & 10.4               & 5.35                 & 4.81                  & 4.08               & 4.95E-01      & 1.02          \\
 PEDOT:PSS        & 10.5               & 5.35                 & 4.78                  & 4.08               & 7.50E-01      & 1.01          \\
 \ch{TiO2}        & 11.2               & 5.39                 & 4.68                  & 4.48               & 2.10E-01      & 1.05          \\
 ZnO              & 10.5               & 4.94                 & 4.35                  & 3.79               & 6.73E-01      & 1.03          \\\bottomrule
 \end{tabular}
 }
\end{table}
\begin{table}[ht]
 \centering
 \caption{The calculated entropy, gzip size, bzip size, XZ size, Fisher information, and fractal dimension of the measured electrical signal of the colloids for a chaotic-like signal input}
 \label{tab:lorenz}
 \resizebox{\textwidth}{!}{
 \begin{tabular}{lSSSSSSS}
 \toprule
 \textbf{Colloid} & {\textbf{Entropy}} & {\textbf{gzip Size}} & {\textbf{bzip2 Size}} & {\textbf{XZ Size}} & {\textbf{FI}} & {\textbf{FD}} \\ \midrule       
 Au               & 12.4               & 4.95                 & 4.41                  & 3.58               & 7.83E-01      & 1.02          \\
 Ferrofluid       & 12.4               & 3.44                 & 2.88                  & 2.62               & 9.47E-01      & 1.03          \\
 \ch{g-C3N4}      & 12.3               & 4.58                 & 3.99                  & 3.29               & 6.32E-01      & 1.06          \\
 MXene            & 12.4               & 4.65                 & 4.18                  & 3.23               & 7.71E-01      & 1.00          \\
 PEDOT:PSS        & 12.4               & 5.10                 & 4.43                  & 3.80               & 8.77E-01      & 1.04          \\
 \ch{TiO2}        & 12.4               & 5.16                 & 4.35                  & 3.99               & 5.60E-01      & 1.07          \\
 ZnO              & 18.5               & 4.43                 & 3.80                  & 3.03               & 5.36E-01      & 1.02          \\  \bottomrule
 \end{tabular}}
\end{table}
 
\begin{table}[ht]
 \centering
 \caption{The calculated entropy, gzip size, bzip size, XZ size, Fisher information, and fractal dimension of the measured electrical signal of the colloids for an 8-bit random binary signal input}
 \label{tab:Bombardment}
 \resizebox{\textwidth}{!}{
 \begin{tabular}{lSSSSSSS}
 \toprule
 Colloid     & {\textbf{Entropy}} & {\textbf{gzip Size}} & {\textbf{bzip2 Size}} & {\textbf{XZ Size}} & {\textbf{FI}} & {\textbf{FD}} \\ \midrule
 Au          & 12.6               & 5.92                 & 5.05                  & 5.32               & 6.16E-02      & 1.08          \\
 Ferrofluid  & 11.3               & 4.18                 & 3.36                  & 3.50               & 9.11E-01      & 1.08          \\
 \ch{g-C3N4} & 20.5               & 5.90                 & 5.07                  & 5.44               & 1.53E-03      & 1.09          \\
 MXene       & 16.1               & 5.76                 & 4.93                  & 5.21               & 7.95E-03      & 1.11          \\
 PEDOT:PSS   & 12.1               & 5.88                 & 5.08                  & 5.23               & 1.04E-01      & 1.08          \\
 \ch{TiO2}   & 38.0               & 5.87                 & 5.05                  & 5.34               & 1.23E-03      & 1.10          \\
 ZnO         & 15.9               & 5.68                 & 4.86                  & 5.11               & 9.75E-03      & 1.09          \\ \bottomrule
 \end{tabular}}
\end{table}

\begin{figure}[hb]
 \centering
 \includegraphics[width=0.775\textwidth]{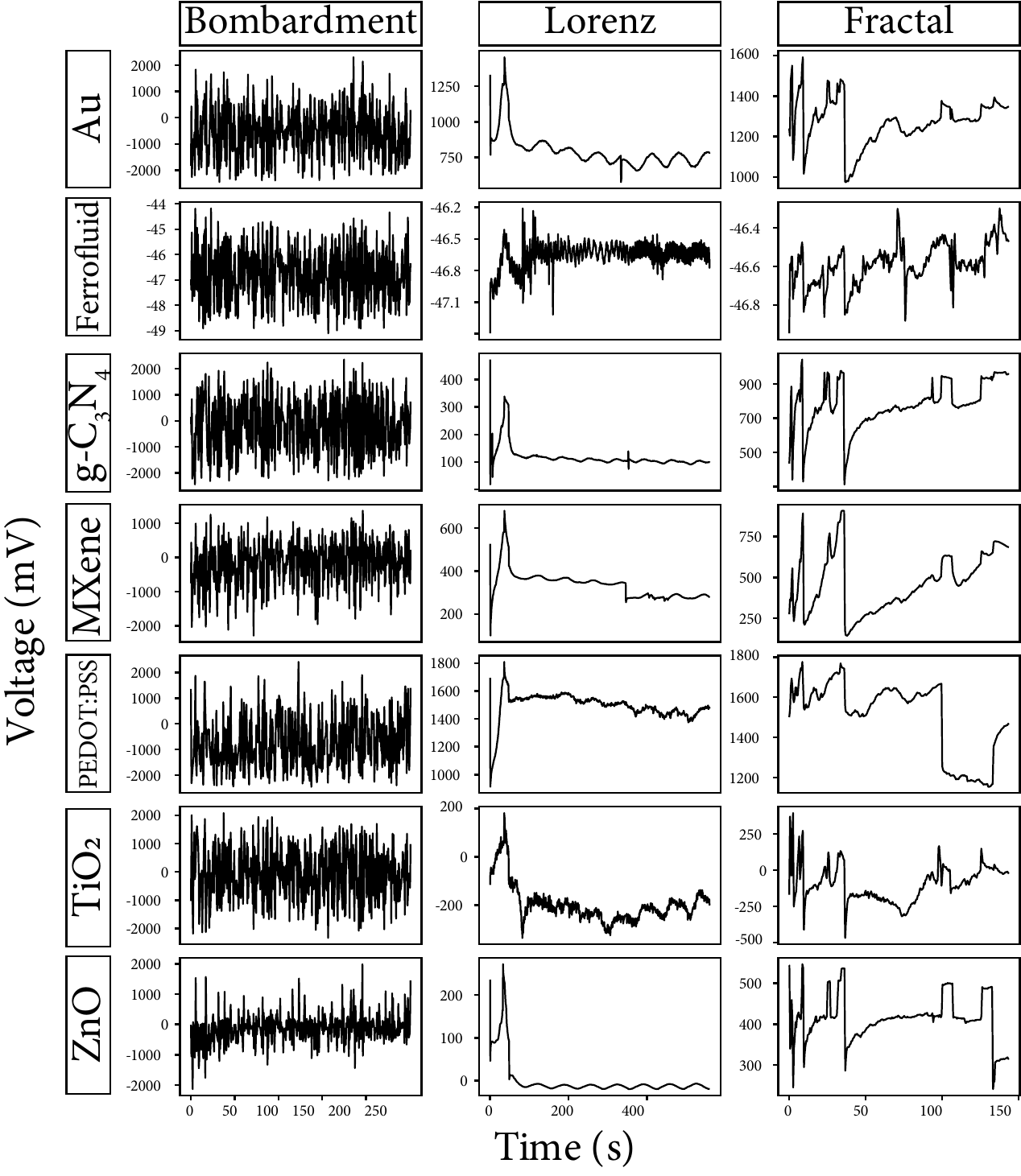}
 \caption{Measured electrical responses of the colloids for random, chaotic, and fractal signal inputs.}
 \label{fig:measured_signal}
\end{figure}

The colloid with the highest complexity across all inputs was \ch{TiO2}. For the fractal input, Au and PEDOT:PSS secured the second and third positions, while for the chaotic input, PEDOT:PSS and Au had the second and third positions, respectively. Finally, for the random binary string input, \ch{g-C3N4} and Au had the second and third positions.
Fig.~\ref{fig:measured_signal} shows the recorded electrical signals of the colloids for all inputs. Visually, the highly complex behavior of the \ch{TiO2} colloid is evident, characterized by an unpredictable pattern.
\FloatBarrier

\begin{figure}[hb]
 \centering
 \begin{subfigure}[b]{0.475\textwidth}
 \caption{}
 \includegraphics[width=\textwidth]{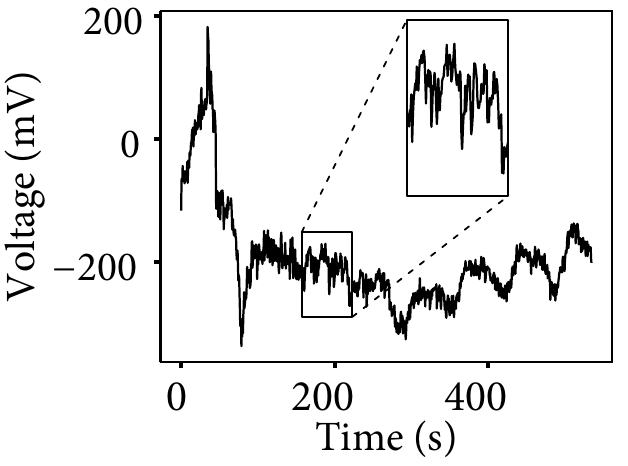}
 \label{fig:tio2}
 \end{subfigure}
 \begin{subfigure}[b]{0.475\textwidth}
 \caption{}
 \includegraphics[width=1\textwidth]{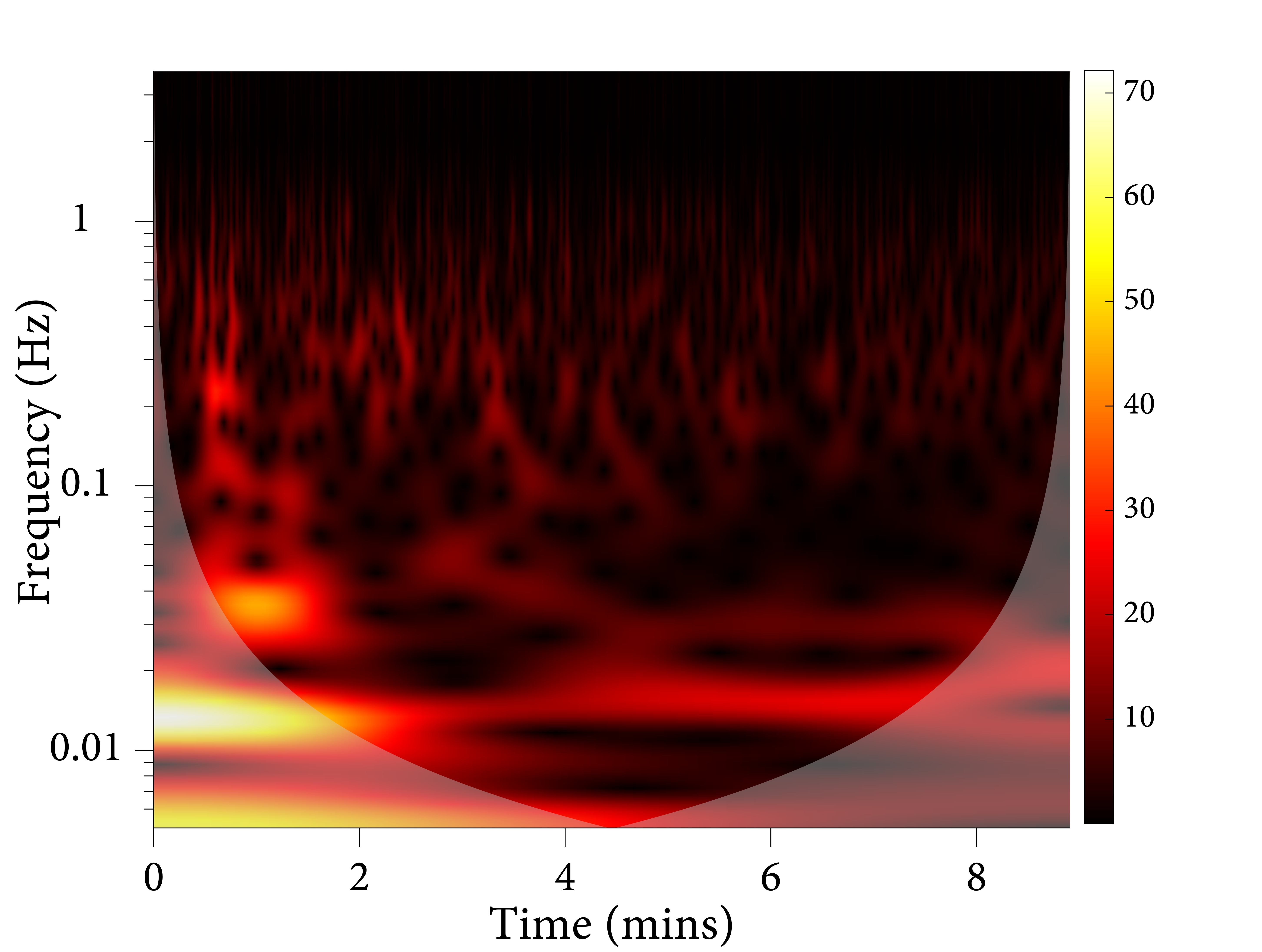}
 \label{fig:scalogram}
 \end{subfigure}
 \caption{(a) Measured electrical response for a chaotic signal for the \ch{TiO2} colloid. The inset shows a zoomed-in region of the curve to elucidate the complex and unpredictable pattern exhibited by the colloid when subjected to electrical signals. (b) Frequency-time wavelet spectrogram of the measured electrical response for a chaotic signal for the \ch{TiO2} colloid. The predominantly dark plot with limited areas of light shading indicates a wide distribution of frequencies over time and a non-repetitive pattern in the data.}
 \label{fig:three graphs}
\end{figure}

Fig.~\ref{fig:tio2} shows the measured electrical response for a chaotic signal input for the \ch{TiO2} colloid. The inset within \ref{fig:tio2} zooms in on a region of the curve, highlighting the complex and unpredictable nature of the colloid's behavior when subjected to electrical signals. In addition, Fig.~\ref{fig:scalogram} shows the frequency-time wavelet spectrogram of the measured electrical response for the chaotic signal applied to the \ch{TiO2} colloid. This spectrogram provides a dynamic representation of the response of the colloid over time, highlighting the distribution of frequencies and their variations. The predominantly dark plot, with limited areas of light shading, indicates a wide dispersion of frequencies over time and a distinct lack of repetitive patterns in the data, highlighting the highly complex and dynamic nature of the response of the \ch{TiO2} colloid and its dynamical behavior under electrical stimulation.

Tab.~\ref{tab:input} shows the indexes calculated for the input electrical signals. As can be seen, the indexes are generally lower for the inputs than what has been observed for all colloid solutions, which indicates an increase in the complexity of the signal.
\begin{table}[ht]
 \centering
 \caption{The calculated entropy, gzip size, bzip size, XZ size, maximal Lyapunov exponent, Fisher information, and fractal dimension of the input electrical signals}
 \label{tab:input}
 \resizebox{\textwidth}{!}{
 \begin{tabular}{lSSSSSSS}
 \toprule
 \textbf{Colloid} & {\textbf{Entropy}} & {\textbf{gzip Size}} & {\textbf{bzip2 Size}} & {\textbf{XZ Size}} & {\textbf{FI}} & {\textbf{FD}} \\ \midrule
 Fractal          & 3.48               & 6.13E-02             & 7.71E-02              & 8.80E-02           & 4.25E-01      & 1.31          \\
 Lorenz           & 4.50               & 1.00E-02             & 1.43E-02              & 2.16E-02           & 5.25E-01      & 1.00          \\
 Random           & 9.99E-01           & 2.38E-01             & 2.53E-01              & 3.56E-01           & 1.28E-01      & 1.98          \\ \bottomrule
 \end{tabular}}
\end{table}

The increase in complexity was likely due to many processes occurring when an electric field is applied to the colloid. Ionizable and charged particles undergo translational movement and transient conduction due to the presence of an applied external field~\cite{bazant2010induced}. Dissociation of neutrals into charged species and recombination of ionized species take place in regions where the charge concentration deviates from equilibrium~\cite{novotnyContributionsParticlesElectrical1986}. Charged species can also induce motion of uncharged neutral liquids, affecting charge transport and creating electrohydrodynamic conditions~\cite{attenElectrohydrodynamicStabilityDielectric1974}. Particle transport can also arise from phenomena like dielectrophoresis~\cite{pishnyakAggregationColloidalParticles2011}, linear and nonlinear electrophoreses~\cite{khairNonlinearElectrophoresisColloidal2022}, and the Quincke rotational effect~\cite{pradilloQuinckeRotorDynamics2019}. Simulating these dynamics is not an easy task~\cite{zhouDynamicDielectricResponse2013,zhouDielectricResponseNanoscopic2012,krishnamurthyElectricResponseTopological2021,duttaTransientDynamicalResponses2018} thus the measurements used here provide a fast way of selecting the most complex colloids without performing time-consuming simulations.
 
The observed increase in entropy along with file compression sizes suggests a potential use of colloidal solutions for reservoir computing. The correlation between entropy and classifier performance is well established in the machine learning community~\cite{ baldanComplexityMeasuresFeatures2023} and serves as a guiding principle for model optimization and performance evaluation~\cite{karacaCharacterizingComplexitySelfSimilarity2020,karacaNOVELFRACTALANALYSIS2020}. The observed trends in entropy and file compression sizes for the colloidal solutions for all inputs not only indicate possible good classification performance but also underscore the potential effectiveness of colloids in processing and extracting meaningful information from time series data.
In addition, the data reveals a consistent trend in which the fractal dimension of the colloidal is generally lower than that of the input, suggesting the existence of memory within the colloidal solution~\cite{Qian2005HURSTEA}. This indicates the possibility of the usage of colloids as \emph{in-materia} processing of time series data for high accuracy predictions~\cite{diaconescuUseNARXNeural2008,ghoshFractalInspectionMachine2018,qianStockMarketPrediction2007,raubitzekFractalInterpolationApproach2021,raubitzekTamingChaosNeural2021}.
\section{Black-box Nonlinearity}
To measure the nonlinearity introduced by the colloid solution in the applied voltage, we define a measure of the nonlinearity. Given an input vector space $V$, an output vector space $W$, and a map $M:V\to W$, the ``amount'' of nonlinearity of the map in a region $\alpha\subseteq V$ is defined as
\begin{equation}
 L^{(M)}_\alpha =   \left\lVert  M(v) - P(\hat{\theta}, v)\right\rVert_{\alpha} = \sqrt{\int_{\alpha} \left [M(v) - P(\hat{\theta}, v)\right ]^2\mathrm{d}v},
 \label{eq:nonlinear_1}
\end{equation}
where $v\in V$ and $P(\hat{\theta}, v)$ is the hyperplane that best fits the data around the region $\alpha$, and it is of the following form
\begin{equation}
 P(\hat{\theta}, v) = \sum_{j} \hat{\theta^j}\cdot v^j +b,
\end{equation}
where $\hat{\theta}^{j} \in \hat{\theta}$ is the $j$-th estimated coefficient of the hyperplane, $v^{j}$ is the $j$-th component of $v$, and $b$ is an offset.
In practice, we are sampling the region $\alpha$ around a discrete lattice, so it is safe to assume a finite number of vectors $n$. Then, the amount of nonlinearity can be defined as
\begin{equation}
 L^{(M)}_{\alpha} = \sqrt{\sum_{i} \left [M(v_i) - P(\hat{\theta}, v_i)\right ]^2},
 \label{eq:nonlinear_2}
\end{equation}
where $\alpha$ is no longer a continuous region in the input space but a collection of input vectors $\lbrace v_i\rbrace, i=1,\dots,n$.
Note that the summation in Eq.~\ref{eq:nonlinear_2} may be interpreted as an approximation of the integral quantity defined in Eq.~\ref{eq:nonlinear_1}. Fundamentally, this nonlinearity measure shows that the output of the mapping $M$ can be described as a linear combination of the input vectors plus a constant. Physically, this offset represents a possible potential present in the suspension when the measurement started.

Tab.~\ref{tab:nonlinearity} presents the estimated nonlinearity of the mapping of each colloidal system from the input stimulus to the measured output signal. Notably, the colloids that exhibited the highest complexity are also estimated to possess the highest amount of nonlinearity.
In the context of reservoir computing, a crucial aspect of the reservoir is its capacity to perform a nonlinear mapping of the input data into a high-dimensional nonlinear state space. This nonlinear transformation of the input data enables the reservoir to capture and encode complex relationships and patterns present within the input data, allowing the subsequent linear readout layer to effectively extract and leverage these encoded features for tasks such as classification, regression, or prediction~\cite{tanakaRecentAdvancesPhysical2019}. The high degree of nonlinearity of the colloids indicates their possibility for inclusion as nonlinear reservoirs in reservoir computing.
 
\begin{table}[]
 \centering
 \caption{Estimated nonlinearity of the output signal of the colloids, based on the input signals}
 \label{tab:nonlinearity}
 \begin{tabular}{lrrr}
 \toprule
 & \multicolumn{3}{c}{$L^{(M)}_{\alpha}$} \\
 Colloid     & Fractal & Lorenz & Bombardment \\ \midrule
 Au          & 1.57    & 2.27   & 13.51       \\
 Ferrofluid  & 0.07    & 0.13   & 0.67        \\
 \ch{g-C3N4} & 2.32    & 2.19   & 15.61       \\
 MXene       & 3.15    & 4.20   & 9.24        \\
 PEDOT:PSS   & 3.54    & 2.56   & 15.43       \\
 \ch{TiO2}   & 7.42    & 8.14   & 14.52       \\
 ZnO         & 3.27    & 3.88   & 7.79        \\ \bottomrule
 \end{tabular}
\end{table}
\FloatBarrier
\section{Conclusions}
This work investigated the complexity and nonlinearity of various colloidal suspensions subjected to different electrical signal inputs—fractal, chaotic, and random binary signals. Several complexity measures were applied to the recorded electrical output signals from the colloids, including entropy, file compression sizes, fractal dimension, and Fisher information. The most complex colloidal suspension across all input signals was found to be \ch{TiO2}, exhibiting high entropy, poor compressibility, and an unpredictable response pattern. The high complexity observed in the \ch{TiO2} and other colloids suggests their potential usage for reservoir computing applications where transforming inputs into a high-dimensional nonlinear state space is desirable.
To quantify the nonlinearity introduced by the colloids, a measure was defined based on the deviation of the output signal from the best-fit hyperplane of the input-output mapping. The results indicated that the colloids exhibiting higher complexity also possessed a higher degree of nonlinearity in their electrical responses.
Overall, this study provides insights into the complex dynamics that can emerge in colloidal suspensions under electrical stimulation. The observed high complexity and nonlinearity highlight the potential of these unconventional computing materials for applications in neuromorphic computing, time series processing, and other domains that can exploit their rich dynamical properties. Further research is warranted to better understand the underlying mechanisms driving these complex phenomena and to explore their computational capabilities in greater depth.
\bibliographystyle{elsarticle-num}
\bibliography{references}
\end{document}

\title{Supporting Information for ``Complexity and nonlinearity of colloid electrical transducers''}
 
\author[1]{Raphael Fortulan\footnote{Corresponding author: raphael.vicentefortulan@uwe.ac.uk}}
\author[1]{Noushin Raeisi Kheirabadi}
\author[2,1]{Alessandro Chiolerio} 
\author[1]{Andrew Adamatzky}
\affiliation[1]{Unconventional Computing Laboratory, University of the West of England, Bristol, UK}
\affiliation[2]{Center for Bioinspired Soft Robotics, Istituto Italiano di Tecnologia,  Genova, Italy}

\startlist{toc}
\begin{abstract}
\vspace{-48pt}
\printlist{toc}{}{\section*{}}
\end{abstract}
\maketitle
\section*{}
\parindent0pt

\section{Koch snowflake}
\begin{algorithm}[H]
 \caption{Koch Snowflake Generator}
 \label{alg:Koch}
 \begin{algorithmic}[1]
        
 \Function{koch}{$x_1, y_1, x_2, y_2, i$}
 \State $angle \gets 60 \times \pi / 180$
 \State $x_3 \gets (2 \times x_1 + x_2) / 3$
 \State $y_3 \gets (2 \times y_1 + y_2) / 3$
 \State $x_4 \gets (x_1 + 2 \times x_2) / 3$
 \State $y_4 \gets (y_1 + 2 \times y_2) / 3$
 \State $x_5 \gets x_3 + (x_4 - x_3) \times \cos(angle) + (y_4 - y_3) \times \sin(angle)$
 \State $y_5 \gets y_3 - (x_4 - x_3) \times \sin(angle) + (y_4 - y_3) \times \cos(angle)$
        
 \If{$i > 0$}
 \State \Call{koch}{$x_1, y_1, x_3, y_3, i - 1$}
 \State \Call{koch}{$x_3, y_3, x, y, i - 1$}
 \State \Call{koch}{$x, y, x_4, y_4, i - 1$}
 \State \Call{koch}{$x_4, y_4, x_2, y_2, i - 1$}
 \Else
 \For{$j=1,3,5,4,2$}
 \State Send to DAC $x_i$, $y_i$
 \State Delay for 100 ms
 \EndFor
 \EndIf
 \EndFunction
 \State \textbf{start:}
 \State Define initial conditions $x_1$, $y_1$, $x_2$, $y_2$
 \For{$i$ from 0 to 4}
 \State \Call{koch}{$x_1$, $y_1$, $x_2$, $y_2$, $i$}
 \EndFor
 \end{algorithmic}
\end{algorithm}
\section{Fractal dimension}
Fractal dimension is a measure of the complexity of a signal that measures the complexity that is being repeated in the signal. For a one-dimensional time series, the fractal dimension can be understood as the space taken by the signal when placed on a uniform 2D grid; in this case, it is the ratio of the log number of boxes $\log (N)$ by the log of the inverse length of the boxes $\log (1/L)$~\cite{wuEffectiveMethodCompute2020}. The fractal dimension is closely related to the persistence of the time series or correlation by the Hurst exponent, and the fractal dimension $D$ of a time series can be evaluated as
\begin{align}
 D = 2 - H. 
\end{align}

Both the Hurst exponent and the fractal dimension measure the roughness of a time series, with a lower $H$ value implying a more rough and therefore complex curve~\cite{merazMultivariateRescaledRange2022,tangComplexityTestingTechniques2015}.
The calculation of the Hurst exponent follows the rescaled range analysis ($R/S$ analysis) as follows: Let $X = [X (t_0), X (t_1), \ldots, X (t_{N - 1})]$ be a univariate time series, then the mean value of $X$ can be expressed as
\begin{align}
 \langle X \rangle = \frac{1}{N} \sum^{N - 1}_{i = 0} X (t_i) 
\end{align}

The cumulative deviation is calculated as
\begin{align}
 \delta (t_i) = \sum^i_{j = 0} X (t_j) - \langle X \rangle, 
\end{align}
for $i = 0, \ldots, N - 1$.

The range $R$ of the time series can then be expressed as
\begin{align}
 R (t_i) = \max_{j = 0, \ldots, i} \delta (t_j) - \min_{k = 0, \ldots, i} 
 \delta (t_k),                         
\end{align}
for $i = 0, \ldots, N - 1$.

The standard deviation series is obtained as
\begin{align}
 S (t_i) = \sqrt{\frac{1}{i + 1} \sum^i_{j = 0} (X (t_j) - \langle X 
 \rangle_i)^2}, i = 0, \ldots, N - 1,       
\end{align}
where $\langle X \rangle_i$ is the average over the time $t_0 \infixand t_i$.

The rescaled range series is then evaluated as
\begin{align}                                                                 
 \left( R / S \right) (t_i) = \frac{R (t_i)}{S (t_i)}, \label{eq:RS}
\end{align}
for $i = 0, \ldots, N - 1$.

The asymptotic behavior of Eq.~\ref{eq:RS} for an independent random process with finite variance follows a power-law function
\begin{align}
 \left( R / S \right) (t_i) = C t_i^H,        
\end{align}
where $C$ is a constant and $H$ is the Hurst exponent.
\section{Entropy}
The entropy of a signal can be thought of as the ``amount of information'' contained in the signal. Therefore, a higher value of entropy can be understood as belonging to a more complex signal.
\subsection{ Shannon's Entropy}
Shannon's entropy is a metric that quantifies the minimum amount of storage required to store and transmit the data~\cite{shannonMathematicalTheoryCommunication1948}. For a given discrete signal, $X = [X_0, X_1, \ldots, X_{N - 1}]$, with a probability of occurrence of each value $P(X_0), \ldots, P (X_{N - 1})$, the Shannon's entropy of $X$ can be defined as
\begin{align}
 H (X) \triangleq - \sum^{N - 1}_{i = 0} P (x_i) \log_2 P (x_i),
\end{align}
where $H (X)$ is given in bits.
\section{Fisher Information}
The Fisher information of a signal is a way of measuring the amount of information that the system carries about itself. Its value is anti-correlated to the other complexity measurement since the more information the system carries about itself, the more predictable it is and, by consequence, less complex.

The first step to extract the Fisher information of a signal is to perform the time-delay embedding of a signal~\cite{dawReviewSymbolicAnalysis2003}. The central idea of time-delay embedding is to reconstruct the phase-space dynamics of a multidimensional system by using observations of a single observable
\begin{align}
 \vec{X} = \{ x_0, x_1, \ldots, x_{N - 1} \} 
\end{align}
by plotting the observations in a phase space of lagged coordinates
\begin{align}
 \vec{\xi} _i = [x_i, x_{i + \tau}, x_{i + 2 \tau}, \ldots, x_{i + (m - 1) 
 \tau}],                        
\end{align}
where $\tau$ is the embedding delay and $m$ is the embedding dimension. The constructed embedded space is then
\begin{align}
 \Xi = [\vec{\xi} _0, \vec{\xi} _1, \ldots, \vec{\xi} _{N - 1 - (m - 1) 
 \tau}]^{\top} .                   
\end{align}

The Fisher information is then defined as~\cite{jamesExtractingMultisourceBrain2003}
\begin{align}
 I = \sum^{m - 1}_{i = 0} \frac{(\overline{\sigma}_{i + 1} -         
 \overline{\sigma}_i)^2}{\overline{\sigma}_i}, \overline{\sigma}_i = 
 \frac{\sigma_i}{\sum _{j = 0}^{m - 1} \sigma_i}                     
\end{align}
where $\sigma_i, i = 0, \ldots, m - 1$ is the $i - \tmop{th}$ singular value of $\Xi$ and $ \overline{\sigma}_i, i = 0, \ldots, m - 1$ is the $i -
\tmop{th}$ normalized singular value of $\Xi$.
\section{Lyapunov Exponent}
The Lyapunov exponent of a dynamical system quantifies the rate of separation of infinitely close trajectories. Formally, let the flow $\phi (t, x_0) \in \mathbb{R}^n$ be the unique of $\dot{x} = f (x)$ with an initial condition $x_0$ and $\delta_0$ be an arbitrary small separation vector, then if
\begin{align}
 \phi (t, x_0 + \delta_0) - \phi (t, x_0) \| \approx \exp (\lambda t) \| 
 \phi (0, x_0 + \delta_0) - \phi (0, x_0) \|,                            
\end{align}
$\lambda$ is called the Lyapunov exponent. A positive value of $\lambda$ can indicate a degree of chaos in the system (for more details see, e.g.~\cite{khalilNonlinearSystems2002}).

For time series data containing a single trajectory, the Lyapunov exponent can also be used to examine complexity~\cite{wolfDeterminingLyapunovExponents1985,mendesDecayDistanceAutocorrelation2019}. In general, a higher value of the Lyapunov exponent corresponds to less predictability, i.e., a higher degree of complexity. The rate of separation can be different for different orientations of the initial separation vector. Thus, there is a spectrum of Lyapunov exponents equal in number to the dimensionality of the phase space, but usually, we are only interested in the maximal Lyapunov exponent, which can be defined as
\begin{align}
 \lambda_{\tmop{MLE}} = \lim_{t \rightarrow \infty} \lim_{d  (t) \rightarrow 
 0} \frac{1}{t} \log \frac{d (t)}{d (0)},                                    
\end{align}
where $d (t) = \| \phi (t, x_0 + \delta_0) - \phi (t, x_0) \|$ is the Euclidian distance between the solutions.

For discrete time series data, the calculation of $\lambda_{\tmop{MLE}}$ requires the construction of the embedded phase space. There are a variety of algorithms to compute the maximum Lyapunov exponent, but in this work, we used the method described in~\cite{rosensteinPracticalMethodCalculating1993}, which we briefly explain in sequence.
First, compute the nearest neighbor $\vec{\xi}_{\widehat{j}}$ of the $j$-th point $\vec{\xi}_j$ by minimizing the Euclidean distance over the phase space
\begin{align}
 d_j (0) = \min_{\vec{\xi} _j}  \| \vec{\xi} _j - \vec{\xi} _j \| . 
\end{align}
Since
\begin{align}
 d_j (i) \approx d_j (0) \exp (\lambda_{\tmop{MLE}} (i \Delta t)) \rightarrow 
 \log d_j (i) \approx \log d_j (0) + \lambda_{\tmop{MLE}} (i \Delta t),
\end{align}
where $\Delta t$ is the sampling time and $j =$0, 1, {\dots}, $i + (m - 1) \tau$, then the maximal Lyapunov exponent can be calculated using a least-squares fit of
\begin{align}
 y (i) = \frac{1}{\Delta t} \langle \log d_j (i) \rangle_j 
\end{align}
\section{Compression Algorithms}
Data compression is the art of reducing the number of bits needed to store or transmit data. For instance, consider Morse code, where each letter of the alphabet is coded as a sequence of dots and dashes. Common letters like ``E'' and ``T'' receive shorter codes, while less common ones like ``X'' and ``Z'' are assigned longer codes. All data compression algorithms consist of at least a model and a coder. The model estimates the probability distribution of the strings, and the coder assigns shorter codes to the most likely symbols.

The idea of using compression algorithms to identify the complexity of time series data is that signals exhibiting more redundancy and repeated patterns will compress to a greater degree, indicating lower complexity. Conversely, highly complex or random signals would be less compressible, resulting in lower compression ratios. In this work, we used state-of-the-art compression algorithms, namely, Gzip, bzip2, and LZMA2, utilizing the XZ format.
\section{Gzip}
GNU Gzip uses the DEFLATE algorithm. Its format specification~\cite{deutschRFC1951DEFLATECompressed1996} defines a lossless compression algorithm that combines the LZ77 algorithm~\cite{zivUniversalAlgorithmSequential1977} for identifying repeated strings with Huffman coding~\cite{huffmanMethodConstructionMinimumRedundancy1952} to represent the compressed data efficiently. The algorithm operates by dividing the input data stream into blocks and compressing each block independently.
\section{bzip2}
The bzip2 algorithm employs a multi-stage process that combines the Burrows-Wheeler block sorting transformation~\cite{burrowsBlocksortingLosslessData1994} with Huffman coding. This approach not only achieves high compression ratios but also offers insights into the underlying structure and patterns present in the data.
\section{XZ}
The XZ file format employs the LZMA2 compression method~\cite{thetukaaniprojectXzFileFormat}, a modern version of the Lempel-Ziv-Markov Chain Algorithm (LZMA). LZMA2 uses a large history buffer of up to 4 GB to identify redundancies and patterns more effectively than simpler LZ77 algorithms. It implements optimal parsing to consider shorter matches when doing so ultimately yields a more compact encoding. LZMA also uses repeated match coding to represent recent matches more efficiently. After a match, it uses literal exclusion to prevent inefficiently coding literals that differ from the predicted byte. At its core, LZMA uses an arithmetic coder with advanced context modeling that captures higher-order patterns in the data.
\bibliographystyle{elsarticle-num}
\bibliography{references}